\documentclass[12pt]{article}
\usepackage{amssymb,epsfig}
\newcommand{\es}{$\mathrm{E_6}$}

\begin{document}

\hspace*{\fill}WUE-ITP-2000-008

\hspace*{\fill}LC-TH-2000-025

\hspace*{\fill}hep-ph/0003272

\vspace{1cm}

\begin{center}
\textbf{\Large Neutralino Production with Polarized Beams
in Extended Supersymmetric Models\footnote{
Contribution to the Proceedings of the ``2nd Joint ECFA/DESY Study on
Physics and Detectors for a Linear Electron-Positron
Collider''.}}

\vspace{10mm}
S.~Hesselbach\footnote{e-mail: hesselb@physik.uni-wuerzburg.de}, 
F.~Franke\footnote{e-mail: fabian@physik.uni-wuerzburg.de}, 
H.~Fraas\footnote{e-mail: fraas@physik.uni-wuerzburg.de}

Institut f\"ur Theoretische Physik, Universit\"at W\"urzburg, D-97074
W\"urzburg, Germany
\end{center}

\vspace{5mm}
\begin{abstract}
We discuss associated neutralino production $e^+e^- \rightarrow
\tilde{\chi}_1^0 \tilde{\chi}_2^0$ with
both beams polarized in the MSSM, NMSSM and an \es\ model. 
It is shown that neutralinos with a large singlino component can be
produced at a high luminosity linear collider 
in a broad parameter range. 
Polarization effects in the extended models are compared over the whole
parameter space where the lightest neutralino is mainly a singlino. 
We explain the complete different behavior of these models in 
some parameter regions, which may help to identify the respective
supersymmetric model.
\end{abstract}

\section{Introduction}

A linear collider with both beams polarized will be an excellent tool
not only to discover supersymmetric particles but also to determine
the underlying SUSY model.
Associated neutralino production 
$e^+e^- \to \tilde{\chi}^0_1 \tilde{\chi}^0_2$ could be the first 
kinematically accessible process that may allow a discrimination between the
Minimal Supersymmetric Standard Model (MSSM) and its nonminimal 
extensions.  
Polarization of both beams will clearly facilitate this challenging task.

An extensive analysis of polarization and spin effects in neutralino
production and subsequent decay including the complete spin
correlations in the MSSM has been performed in \cite{gudi}.
This study also includes extended SUSY models:  
the Next-to-Minimal Supersymmetric Standard Model (NMSSM)
with an additional Higgs
singlet superfield and an \es\ inspired model with a Higgs singlet and a new
$\mathrm{U(1)}'$
gauge boson. 
Significant differences between the MSSM and SUSY models with gauge singlets
arise by the neutralino singlet components which do not couple to
(scalar) fermions and standard gauge bosons.
Within a special scenario,
where the masses of the light neutralinos were fixed in the three SUSY
models and the lightest neutralino had a large singlino component in
the extended models, the cross sections for 
neutralino production at an $e^+e^-$ linear collider
with polarized beams have been discussed in \cite{sitges}, and
complete different polarization asymmetries have been found.

Now the study of the polarization effects is extended to a large
region of the parameter space where the lightest 
neutralino is mainly a singlino. 
Comparing the longitudinally polarized cross sections 
we identify the parameter regions where 
polarization effects may help to specify
nonminimal supersymmetric models. Of course for an exact determination
of the SUSY parameters and discrimination between SUSY models a precise
analysis also of the decay characteristics is indispensable. For
a special scenario this was done in \cite{sitges}, the extension to the whole
parameter space is planned.

In the following sections 2 and 3 we introduce the considered
SUSY models
and show the unpolarized cross section for the
neutralino production. Crucial for the understanding
of the polarization effects are the couplings of the left and right
selectrons to the produced neutralinos 
that will be discussed in section 4. A comprehensive comparison of
the longitudinally polarized cross sections in the MSSM and the
extended models in section 5 concludes our contribution.

\section{Extended SUSY: NMSSM and E$_\mathbf{6}$ model}
We shortly present the parameters and the neutralino sectors of the
supersymmetric models.
In the MSSM the neutralino sector depends on the 
gaugino mass parameters $M_2$ and $M_1$, the higgsino mass parameter 
$\mu$ and the ratio of the Higgs vacuum expectation values 
$\tan\beta = v_2/v_1$.
In this paper we assume the GUT-relation $M_1/M_2=5/3 \tan^2 \theta_W$
and scan over the $M$-$\mu$ plane with fixed $\tan\beta =3$.
The longitudinally polarized cross sections for neutralino production are compared with 
two extended SUSY models.

The NMSSM is the simplest extension of the MSSM by a Higgs singlet
field with vacuum expectation value $x$ and hypercharge $0$ \cite{nmssm}.
The masses and couplings of the five neutralinos depend on
$M_2$, $M_1$, $\tan\beta$, $x$ and the trilinear couplings 
$\lambda$ and $\kappa$ in the superpotential. Within the NMSSM, 
a light singlino-like neutralino cannot be experimentally excluded by
LEP \cite{franke}. In order to obtain a light 
singlino-like neutralino $\tilde{\chi}_1^0$ we choose in the following
$x=1000$ GeV and $\kappa=0.01$. Then 
nearly independently of the other parameters the 
$\tilde{\chi}_1^0$ has singlino
character and a mass between about 9 and 28 GeV 
while the masses and mixings of the heavier neutralinos correspond
to the MSSM with $\mu=\lambda x$.

Models with additional U(1) factors in the gauge group are
a further extension of the MSSM beyond the NMSSM. They can be 
motivated by superstring theory \cite{hr} and imply 
an \es\ group as effective
GUT group, which is broken to  
an effective low energy gauge
group of rank 5
with one additional $\rm U(1)'$ factor.
This \es\ model contains one new gauge boson $Z'$ and an
extended Higgs sector with one singlet superfield 
with vacuum expectation value $x$ as in the NMSSM \cite{e6model}.
To respect the experimental mass bounds of new gauge bosons~\cite{abe}
the vacuum expectation value of the singlet must be 
larger than 1~TeV.

The extended neutralino sector in the \es\ model contains six
neutralinos, one $\tilde{Z}'$ gaugino and one singlino $\tilde{N}$
in addition to
the MSSM \cite{e6model,e6neutralino}. The $6 \times 6$ neutralino mass matrix
depends on six parameters: the $\rm SU(2)_L$, U(1)$_Y$ and
$\rm U(1)'$ gaugino 
mass parameters $M_2$, $M_1$ and $M'$, $\tan\beta$, $x$
and the trilinear coupling $\lambda$ in the superpotential.
With the GUT relation $M' = M_1$ between the U(1) gaugino mass
parameters the four lighter neutralinos always have MSSM-like character if
the MSSM parameter $\mu$ is identified with $\lambda x$ \cite{lc97,hesselb}, 
and the two heaviest neutralinos have large
$\tilde{Z}'$ and $\tilde{N}$ components. 
Assuming $|M'| \gg x, M_1$, however,
the lightest neutralino can be a nearly
pure singlino  
\cite{decarlosespinosa}. Then
$\tilde{\chi}^0_2,\dots,  
\tilde{\chi}^0_5$ have MSSM-like character and the $\tilde{\chi}^0_6
\approx \tilde{Z}'$ has a mass ${\cal O}(M')$. Such scenarios
where the spectrum of the lighter neutralinos
is similar to the NMSSM will be discussed in the following.

Table~\ref{scentab} gives an overview over the described fixed parameters,
while the remaining parameters $M_2$ and $\mu$ (or $\lambda x$) were scanned
in the region $0 \le M_2 \le 500$ GeV, $-500 \le \mu\;(\lambda x) \le 500$ GeV
observing the experimental bounds from the unsuccessful chargino and
neutralino search at LEP2 \cite{lepbounds}. Also shown are the masses and
singlino components of the lightest neutralino in the extended models which do
not extensively vary over the scanned parameter region. 
The selectron masses $m_{\tilde{e}_L} = 176$~GeV and
$m_{\tilde{e}_R} = 132$~GeV are motivated
by the DESY/ECFA reference scenario \cite{refszen} and are fixed
in all three models, for comparison.

\begin{table}[ht]
\begin{center}
\renewcommand{\arraystretch}{1.3}
\begin{tabular}{|c|c|c|c|c|c|c|c|c|c|}
\hline
Model & $M'/$TeV & $x/$TeV & $\kappa$ & $\tan\beta$ &
$m_{\tilde{\chi}_1^0}/$GeV & $ |\langle\chi_1^0|\Psi_N\rangle|^2$\\ 
\hline\hline
MSSM & --- & --- & ---  & 3 & 49.3 -- 249.4& 0 \\ \hline
NMSSM & --- & 1 & 0.01 & 3 & 8.7 -- 27.8 & 0.905 -- 0.971 \\ \hline
\es & $-50$ & 3 & --- & 3 & 30.5 -- 32.9 & 0.979 -- 0.996\\ \hline
\es & $+50$ & 3 & --- & 3 & 30.9 -- 33.1 & 0.994 -- 0.997\\ \hline
\es & $+35$ & 3 & --- & 3 & 44.6 -- 46.7 & 0.994 -- 0.996\\ \hline
\end{tabular}
\end{center}
\caption{\label{scentab}Parameters of the supersymmetric models. 
  In the extended models the $\tilde{\chi}^0_1$ has dominant singlino
  component $\langle\chi_1^0|\Psi_N\rangle$ in the whole 
  $M_2$-$\lambda x$ parameter space. 
  The selectron masses $m_{\tilde{e}_L} = 176$~GeV and
  $m_{\tilde{e}_R} = 132$~GeV are fixed in all models.}
\end{table}

\section{Unpolarized cross sections}
In this section we address the question if a neutralino with a dominant
singlino component can be directly produced with a sufficient cross
section at a linear collider.

\begin{figure}[ht!]
\centering
\begin{picture}(16,8.5)
  \put(0,.5){\epsfig{file=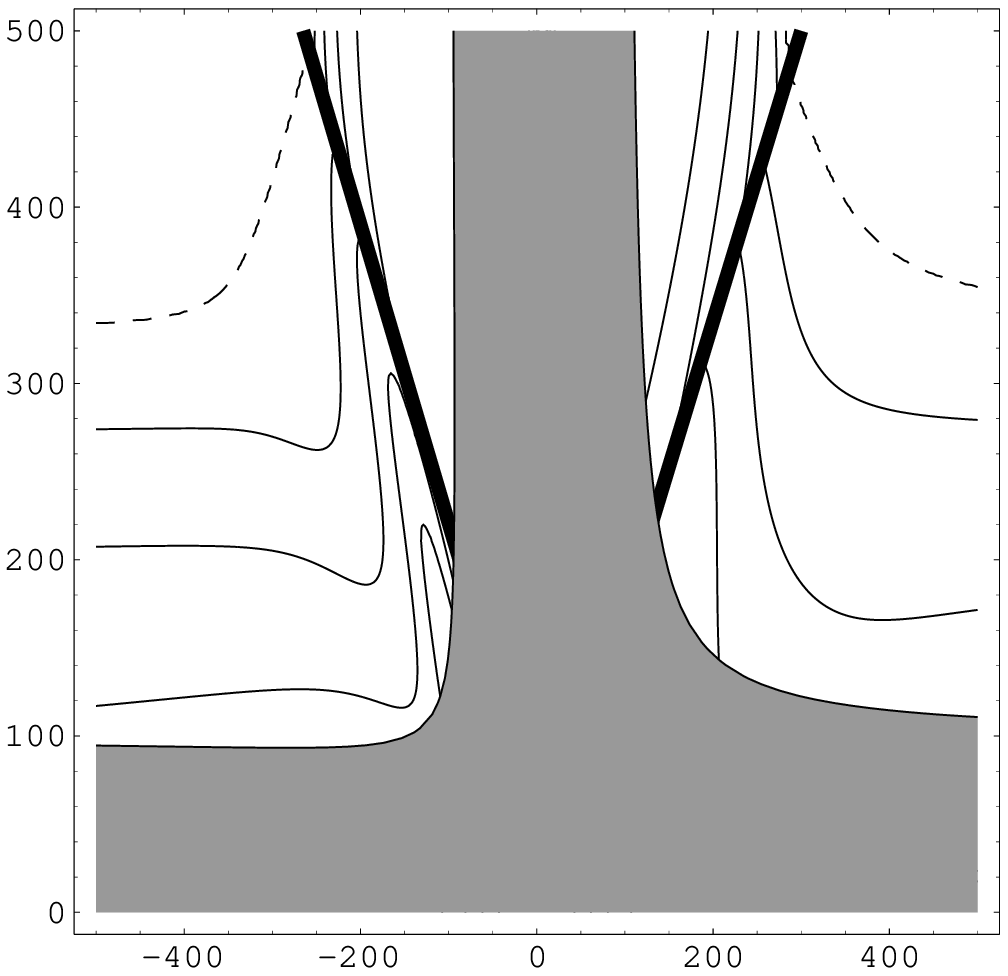,scale=.75}}
  \put(3.5,.1){\small $\mu/$GeV}
  \put(0,8.1){\small $M_2/$GeV}
  \put(3.85,8.1){(a)}
  \put(.7,2.75){\scriptsize 120 fb} \put(5.55,3){\scriptsize 120 fb}
  \put(.7,3.9){\scriptsize 80 fb} \put(6.8,3.4){\scriptsize 80 fb}
  \put(.7,4.8){\scriptsize 40 fb} \put(6.8,4.9){\scriptsize 40 fb}
  \put(2.8,7.5){\scriptsize 160} \put(4.9,7.5){\scriptsize 160}
  \put(2.9,7.2){\scriptsize fb} \put(5.0,7.2){\scriptsize fb}
  \put(8.3,.5){\epsfig{file=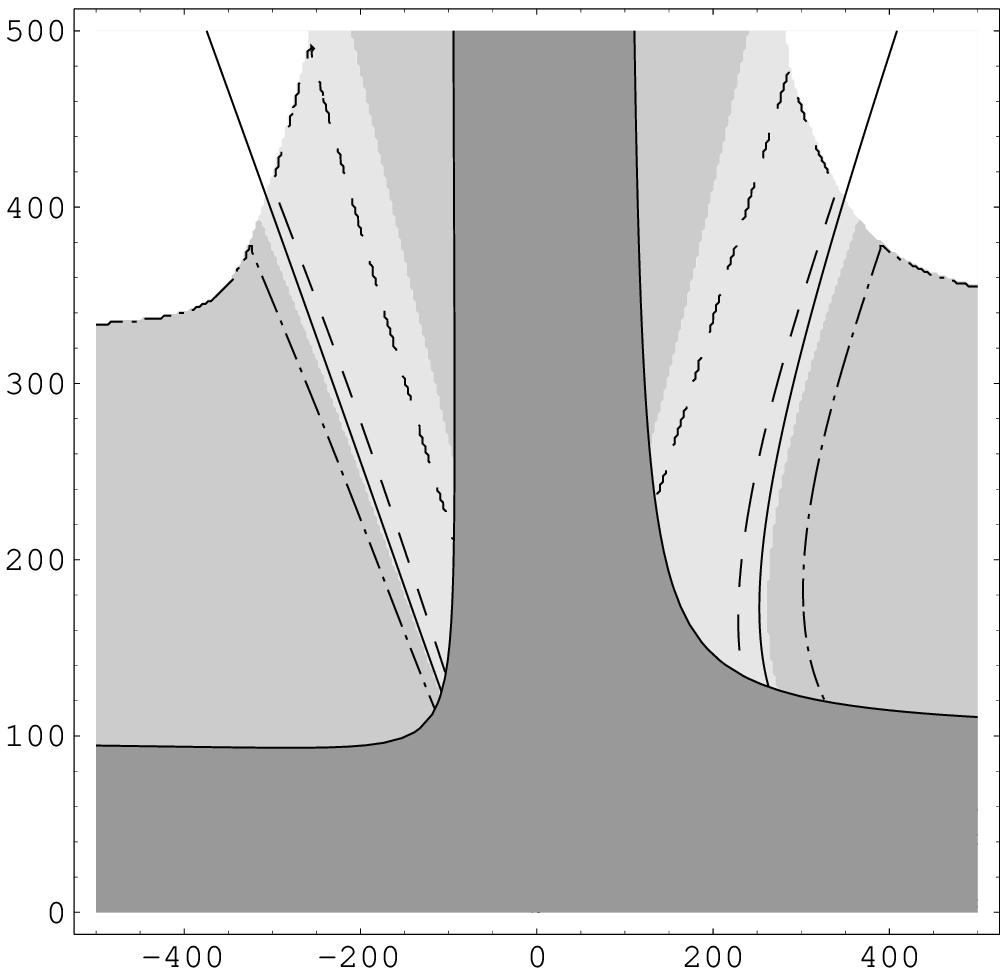,scale=.75}}
  \put(11.8,.1){\small $\mu/$GeV}
  \put(8.3,8.1){\small $M_2/$GeV}
  \put(12.1,8.1){(b)}
  \put(10.7,6){\footnotemark[1]}
  \put(14.2,6){\footnotemark[1]}
  \put(9.7,3.5){\footnotemark[7]}
  \put(15,3.5){\footnotemark[7]}
\end{picture}
\caption[]{\label{mssm}(a) Contour lines 
  of the cross section of
  $e^+e^- \to \tilde{\chi}^0_1 \tilde{\chi}^0_2$ for unpolarized beams 
  in the MSSM for $\sqrt{s}=500$~GeV and $\tan\beta = 3$. 
  (b) Areas where the cross section $\sigma^{(+-)}$ 
  (right handed polarized electrons and left handed polarized positrons,
  bright shaded) or
  $\sigma^{(-+)}$ 
  (left handed polarized electrons and right handed polarized positrons,
  gray shaded) is largest, respectively. 
  In the regions $\ast$ ($\ast\ast$) within the dashed (dashed dotted) 
  lines the ratio $\sigma^{(+-)}/\sigma^{(00)}$
  ($\sigma^{(-+)}/\sigma^{(00)}$) is larger than 2 for $P^-=+0.85$,
  $P^+=-0.6$ ($P^-=-0.85$, $P^+=+0.6$). The solid line
  shows the contour $(f^R_{e1}f^R_{e2})^2/(f^L_{e1}f^L_{e2})^2 = 1$.
  The dark region marks the parameter space excluded by LEP2.}
\end{figure}

The neutralino production cross section in the MSSM is well known
\cite{neutprod}.
Fig.~\ref{mssm}(a) shows the contour lines for the unpolarized
cross section $\sigma^{(00)}$ of $e^+e^- \to
\tilde{\chi}^0_1 \tilde{\chi}^0_2$ at a c.m.s. energy $\sqrt{s}=500$ GeV
in the $M_2$-$\mu$ parameter space. Depending on the masses and mixing
characters of the light neutralinos it takes values up to 200~fb
over most of the parameter space. Note the discontinuities 
(bold lines) for both positive
and negative $\mu$ where $m_{\tilde{\chi}_2^0} = m_{\tilde{\chi}_3^0}$ 
and therefore the mixing character of $\tilde{\chi}_2^0$ changes 
\cite{bartlfraasneut}.

\begin{figure}[ht!]
\centering
\begin{picture}(16,8.5)
  \put(0,.5){\epsfig{file=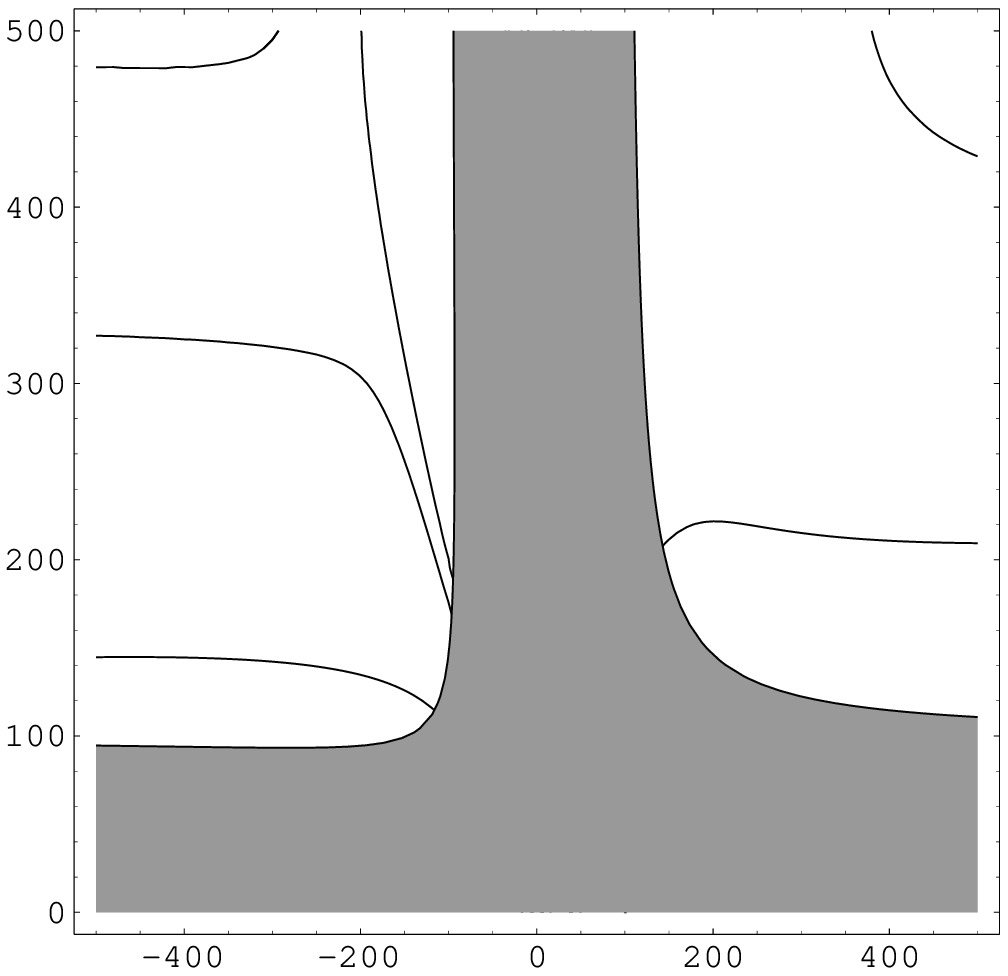,scale=.75}}
  \put(3.5,.1){\small $\lambda x/$GeV}
  \put(0,8.1){\small $M_2/$GeV}
  \put(3.85,8.1){(a)}
  \put(.7,3.05){\scriptsize 5 fb} \put(7.0,3.95){\scriptsize 5 fb}
  \put(.7,5.5){\scriptsize 1 fb} \put(6.95,7.2){\scriptsize 1 fb}
  \put(0.7,7.15){\scriptsize 0.5 fb}

  \put(8.3,.5){\epsfig{file=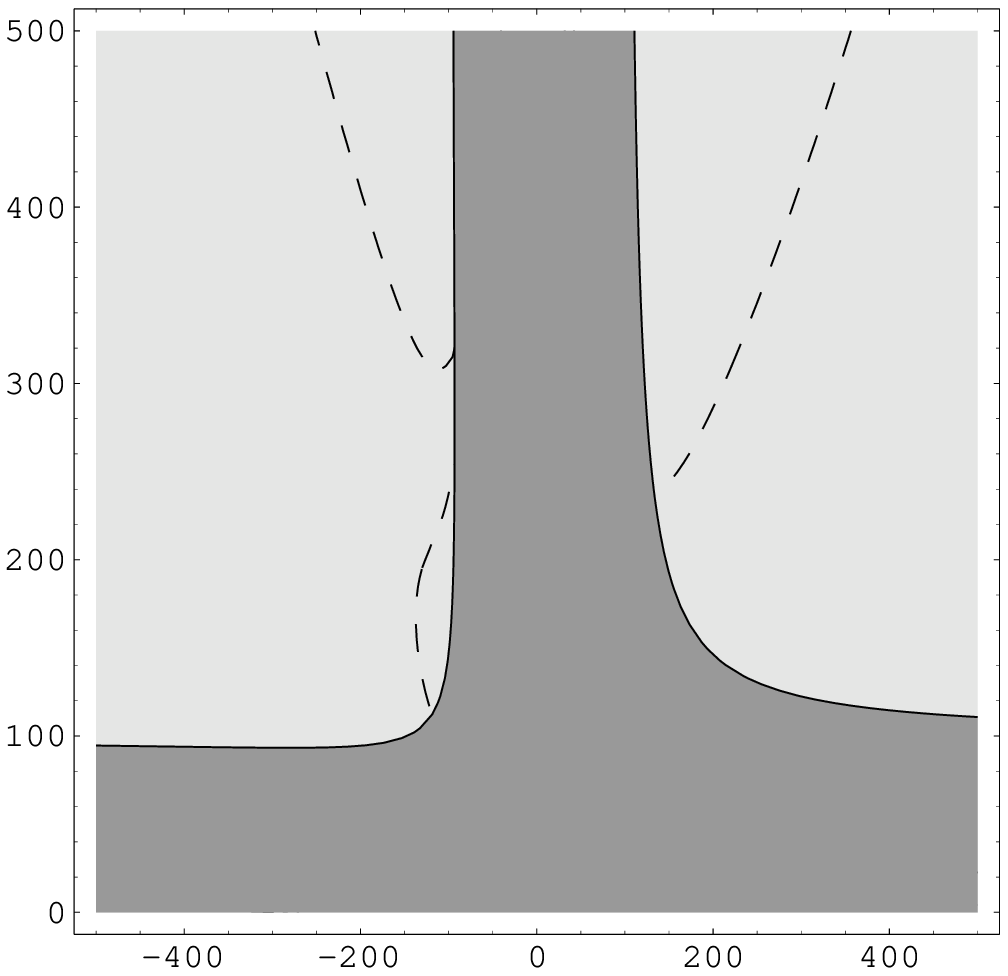,scale=.75}}
  \put(11.8,.1){\small $\lambda x/$GeV}
  \put(8.3,8.1){\small $M_2/$GeV}
  \put(12.1,8.1){(b)}
  \put(9.5,4.5){\scriptsize $\frac{\sigma^{(+-)}}{\sigma^{(00)}} > 2.7$}
  \put(13.9,3.7){\scriptsize $\frac{\sigma^{(+-)}}{\sigma^{(00)}} > 2.7$}

\end{picture}
\caption{\label{nmssm}(a) Contour lines 
  of the cross section of 
  $e^+e^- \to \tilde{\chi}^0_1 \tilde{\chi}^0_2$ for unpolarized beams 
  in the NMSSM with $\kappa = 0.01$ for $\sqrt{s}=500$~GeV and
  $\tan\beta = 3$. (b) In the whole parameter space $\sigma^{(+-)}$
  (right handed polarized electrons and left handed polarized positrons,
  bright shaded) is largest. The dashed lines denote the contours
  $\sigma^{(+-)}/\sigma^{(00)} = 2.7$ for $P^-=+0.85$,
  $P^+=-0.6$. The dark region
  marks the parameter space excluded by LEP2.}
\end{figure}

In the NMSSM with a light singlino-dominated neutralino $\tilde{\chi}^0_1$
the unpolarized cross section (Fig.~\ref{nmssm}(a))
is significantly smaller since the
singlino component does not couple to the $Z$ boson and the selectrons.
It is only the small contribution from
the other components of $\tilde{\chi}_1^0$ that accounts for the cross 
section. Nevertheless one gets even for a  
neutralino  with a singlino component larger than 90\,\%
production
cross sections up to 30~fb which are expected to be clearly visible at
a high luminosity linear collider ($\int {\cal L}=500/$fb) \cite{lumi}.
The use of polarized beams
can still enhance these cross sections as we discuss in the following sections.

\begin{figure}[ht!]
\centering
\begin{picture}(16,8.5)
  \put(0,.5){\epsfig{file=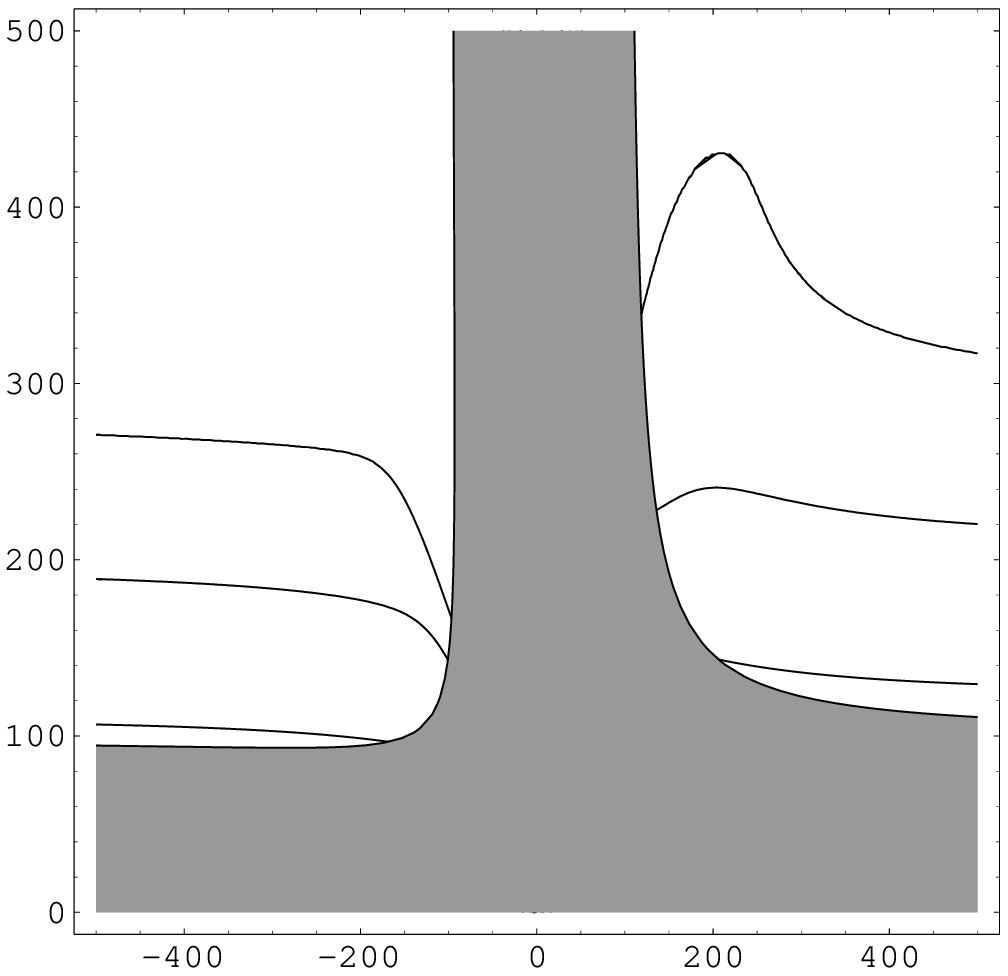,scale=.75}}
  \put(3.5,.1){\small $\lambda x/$GeV}
  \put(0,8.1){\small $M_2/$GeV}
  \put(3.85,8.1){(a)}
  \put(.7,2.55){\scriptsize 5 fb} \put(7.0,2.85){\scriptsize 5 fb}
  \put(.7,3.65){\scriptsize 1 fb} \put(7.0,4.1){\scriptsize 1 fb}
  \put(.7,4.75){\scriptsize 0.5 fb} \put(6.75,5.55){\scriptsize 0.5 fb}

  \put(8.3,.5){\epsfig{file=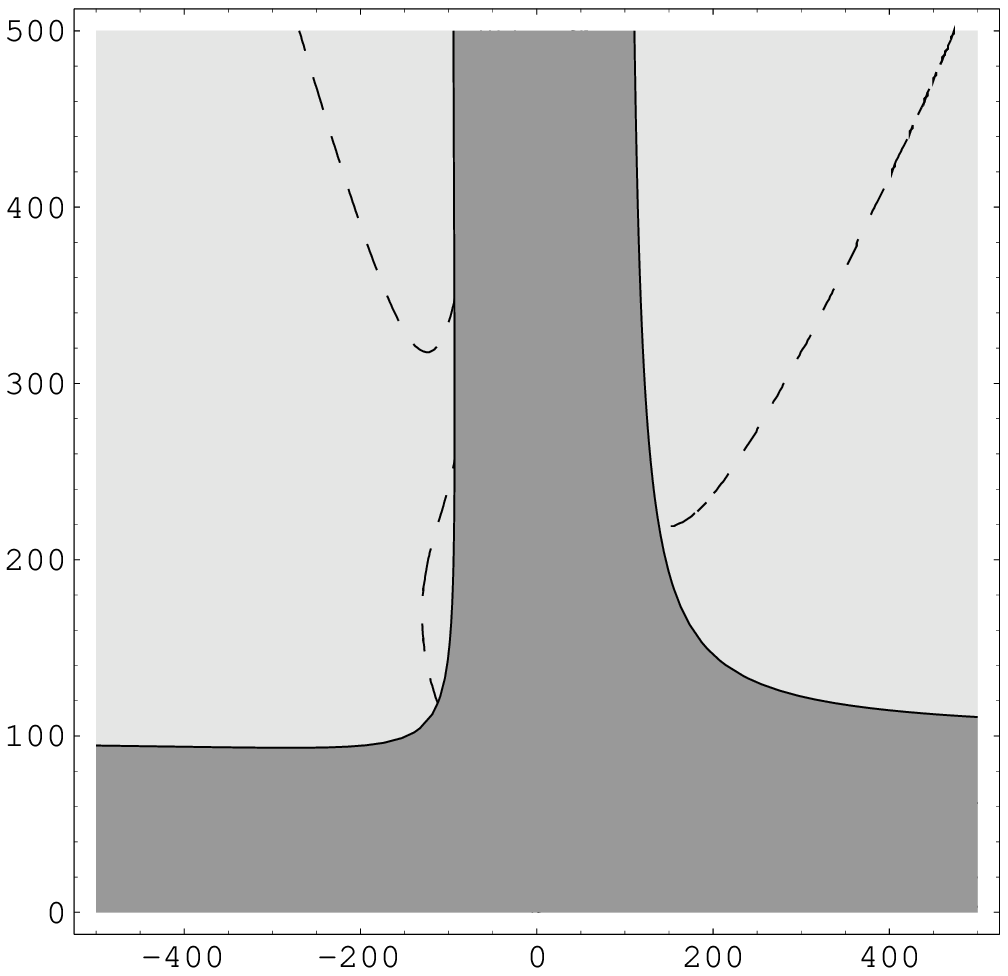,scale=.75}}
  \put(11.8,.1){\small $\lambda x/$GeV}
  \put(8.3,8.1){\small $M_2/$GeV}
  \put(12.1,8.1){(b)}
  \put(9.5,4.5){\scriptsize $\frac{\sigma^{(+-)}}{\sigma^{(00)}} > 2.9$}
  \put(13.9,3.7){\scriptsize $\frac{\sigma^{(+-)}}{\sigma^{(00)}} > 2.9$}

\end{picture}
\caption{\label{e6minus}(a) Contour lines 
  of the cross section of 
  $e^+e^- \to \tilde{\chi}^0_1 \tilde{\chi}^0_2$ for unpolarized beams 
  in the \es\ model with $M' = - 50$~TeV for $\sqrt{s}=500$~GeV and
  $\tan\beta = 3$. (b) In the whole parameter space $\sigma^{(+-)}$
  (right handed polarized electrons and left handed polarized positrons,
  bright shaded) is largest. The dashed lines denote the contours
  $\sigma^{(+-)}/\sigma^{(00)} = 2.9$ for $P^-=+0.85$,
  $P^+=-0.6$. The dark region
  marks the parameter space excluded by LEP2.}
\end{figure}

\begin{figure}[ht!]
\centering
\begin{picture}(16,8.5)
  \put(0,.5){\epsfig{file=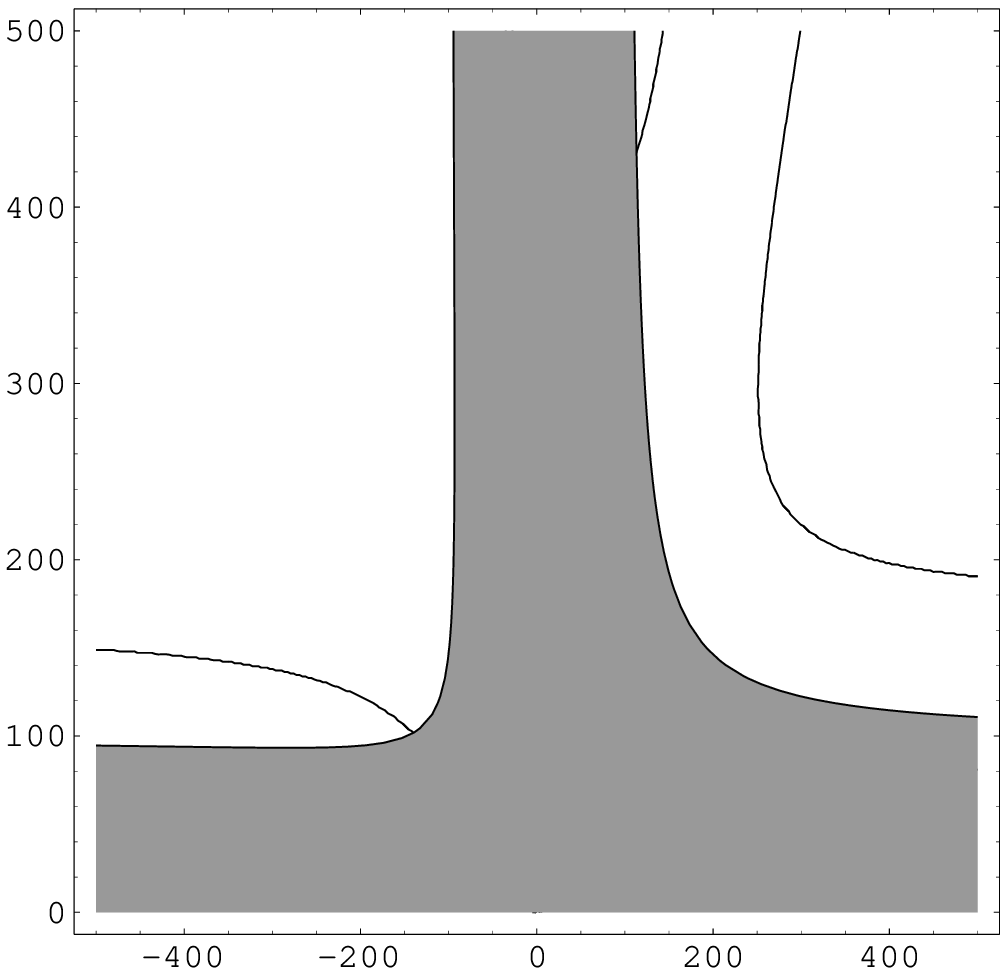,scale=.75}}
  \put(3.5,.1){\small $\lambda x/$GeV}
  \put(0,8.1){\small $M_2/$GeV}
  \put(3.85,8.1){(a)}
  \put(.7,3.15){\scriptsize 0.1 fb}
  \put(5.15,7.5){\scriptsize 0.5 fb}
  \put(6.2,7.5){\scriptsize 0.1 fb}

  \put(8.3,.5){\epsfig{file=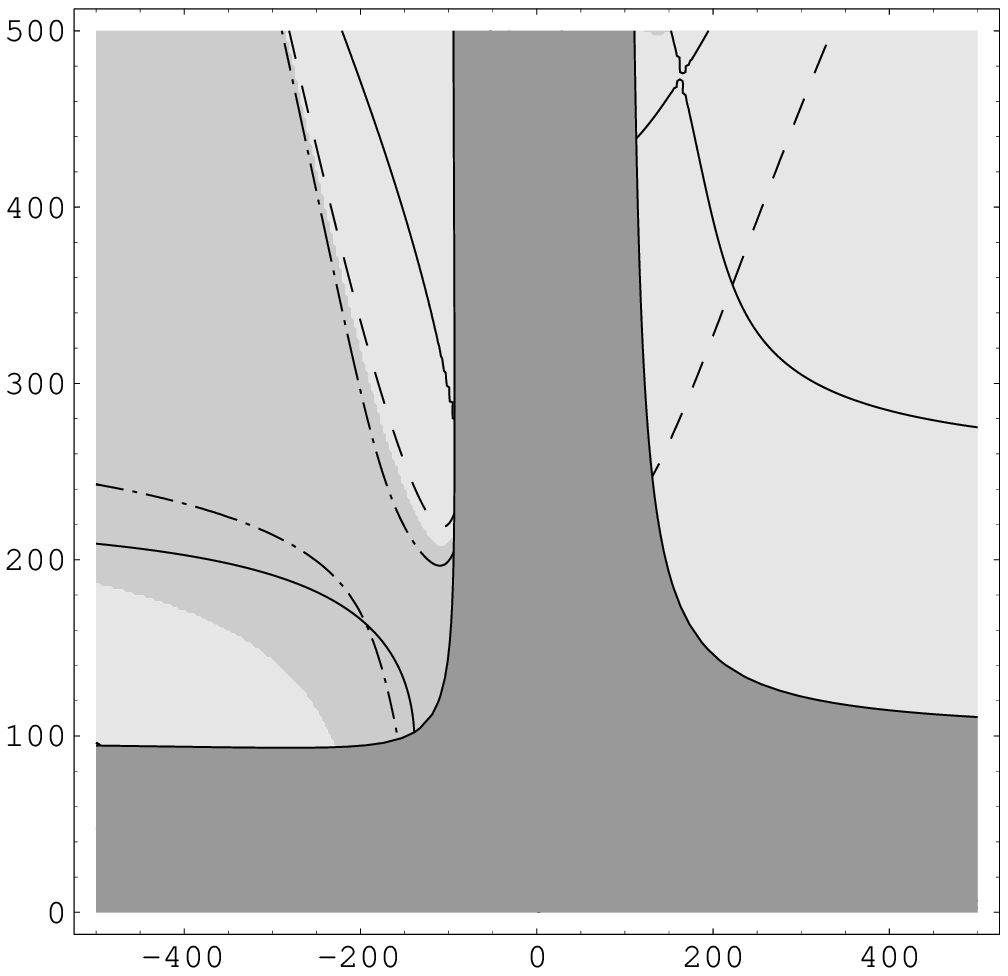,scale=.75}}
  \put(11.8,.1){\small $\lambda x/$GeV}
  \put(8.3,8.1){\small $M_2/$GeV}
  \put(12.1,8.1){(b)}
  \put(14.5,4.5){\footnotemark[1]}
  \put(11.2,7){\footnotemark[1]}
  \put(9.8,5.5){\footnotemark[7]}
\end{picture}
\caption[]{\label{e6plus}(a) Contour lines 
  of the cross section of 
  $e^+e^- \to \tilde{\chi}^0_1 \tilde{\chi}^0_2$ for unpolarized beams 
  in the \es\ model with $M' = + 50$~TeV for $\sqrt{s}=500$~GeV and
  $\tan\beta = 3$. (b) Areas where the cross section $\sigma^{(+-)}$
  (right handed polarized electrons and left handed polarized positrons,
  bright shaded) or $\sigma^{(-+)}$ 
  (left handed polarized electrons and right handed polarized positrons,
  gray shaded) is largest, respectively.
  In the regions $\ast$ ($\ast\ast$) within the dashed (dashed dotted) 
  lines the ratio $\sigma^{(+-)}/\sigma^{(00)}$
  ($\sigma^{(-+)}/\sigma^{(00)}$) is larger than 2 for $P^-=+0.85$,
  $P^+=-0.6$ ($P^-=-0.85$, $P^+=+0.6$).
  The solid line shows
  the contour $(f^R_{e1}f^R_{e2})^2/(f^L_{e1}f^L_{e2})^2 = 1$.
  The dark region marks the parameter space excluded by LEP2.}
\end{figure}

\begin{figure}[ht!]
\centering
\begin{picture}(16,8.5)
  \put(0,.5){\epsfig{file=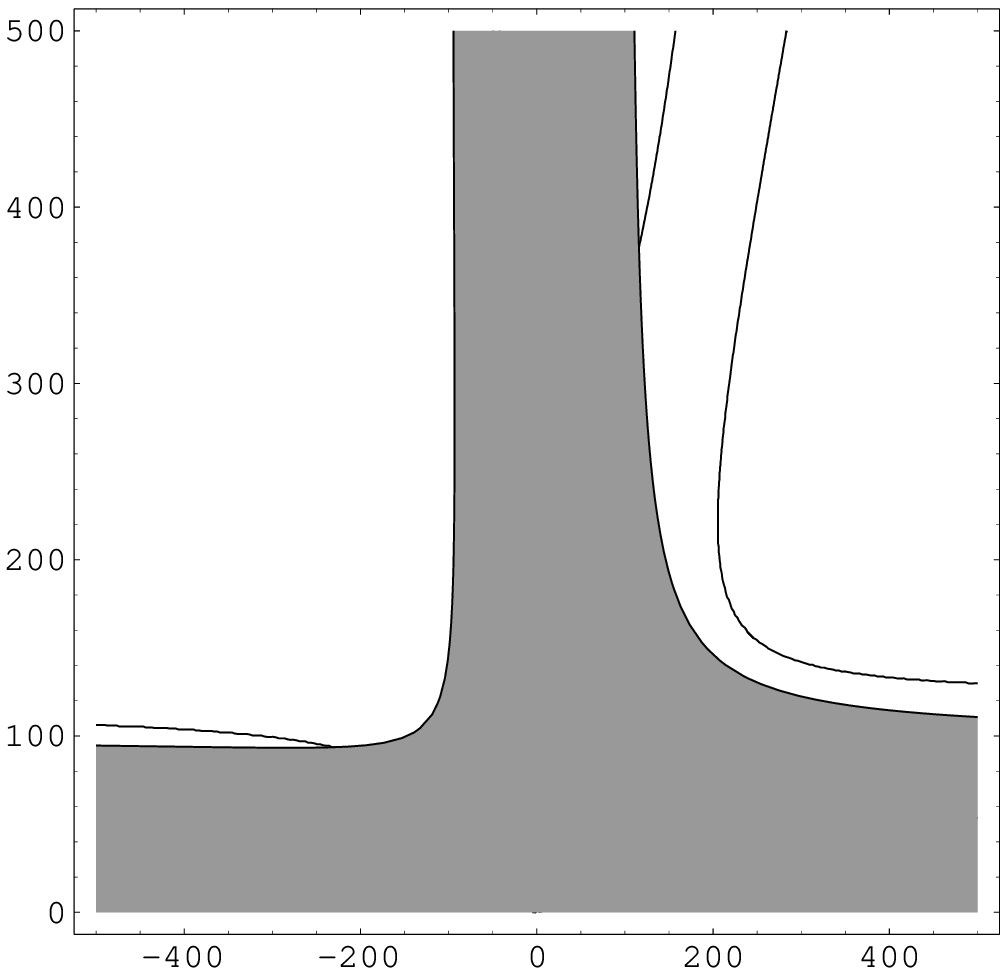,scale=.75}}
  \put(3.5,.1){\small $\lambda x/$GeV}
  \put(0,8.1){\small $M_2/$GeV}
  \put(3.85,8.1){(a)}
  \put(.7,2.55){\scriptsize 0.1 fb}
  \put(5.18,7.5){\scriptsize 0.5 fb}
  \put(6.05,7.5){\scriptsize 0.1 fb}
  \put(8.3,.5){\epsfig{file=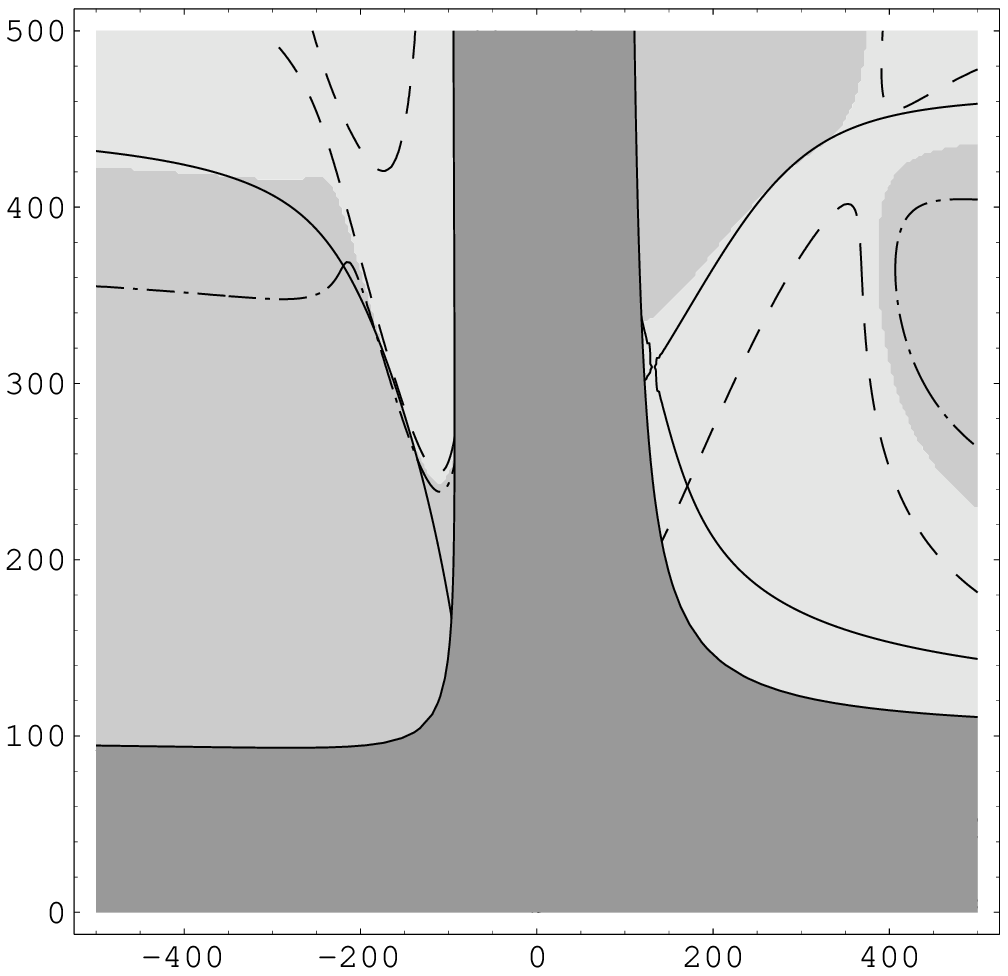,scale=.75}}
  \put(11.8,.1){\small $\lambda x/$GeV}
  \put(8.3,8.1){\small $M_2/$GeV}
  \put(12.1,8.1){(b)}
  \put(11.3,6){\footnotemark[1]}
  \put(14.3,4){\footnotemark[1]}
  \put(15.3,7.3){\footnotemark[1]}
  \put(10,4){\footnotemark[7]}
  \put(15.4,5.6){\footnotemark[7]}
\end{picture}
\caption[]{\label{e6plus35}(a) Contour lines 
  of the cross section of 
  $e^+e^- \to \tilde{\chi}^0_1 \tilde{\chi}^0_2$ for unpolarized beams 
  in the \es\ model with $M' = + 35$~TeV for $\sqrt{s}=500$~GeV and
  $\tan\beta = 3$. (b) Areas where the cross section $\sigma^{(+-)}$
  (right handed polarized electrons and left handed polarized positrons,
  bright shaded) or $\sigma^{(-+)}$ 
  (left handed polarized electrons and right handed polarized positrons,
  gray shaded) is largest, respectively.
  In the regions $\ast$ ($\ast\ast$) within the dashed (dashed dotted) 
  lines the ratio $\sigma^{(+-)}/\sigma^{(00)}$
  ($\sigma^{(-+)}/\sigma^{(00)}$) is larger than 2 for $P^-=+0.85$,
  $P^+=-0.6$ ($P^-=-0.85$, $P^+=+0.6$).
  The solid line shows
  the contour $(f^R_{e1}f^R_{e2})^2/(f^L_{e1}f^L_{e2})^2 = 1$.
  The dark region marks the parameter space excluded by LEP2.}
\end{figure}

In the \es\ model we explicitly study the dependence of the unpolarized
cross sections on the parameter $M'$ in Figs.~\ref{e6minus}(a) --
\ref{e6plus35}(a). For the large negative value 
$M'=-50$ TeV (Fig.~\ref{e6minus}(a)) the masses and the mixings
of the light neutralinos 
are similar as in the NMSSM. Therefore one also obtains similar cross
sections up to 9~fb. For positive $M'$
(Figs.~\ref{e6plus}(a), \ref{e6plus35}(a)), however, the
singlino component is larger than $99.4\,\%$ in the whole parameter
space. Thus the cross sections are reduced 
to maximum values 
of about 0.2~fb for $\lambda x < 0$ and 0.5~fb for
$\lambda x > 0$ leading to at least 100 respective 250 events
at the expected high luminosity linear collider.

\section{Neutralino couplings to selectrons and electrons}
If apart from the singlino component which does not contribute to the
production process both produced neutralinos $\tilde{\chi}_1^0$ and 
$\tilde{\chi}_2^0$ have dominant gaugino components,  
the contribution of $Z$ exchange can be neglected for energies above
the $Z$ peak.
Then for longitudinally polarized beams the cross section consists of
two terms describing the exchange of left and right selectrons $\sigma 
= \sigma_{\tilde{e}_L} + \sigma_{\tilde{e}_R}$. The structure of
$\sigma_{\tilde{e}_{L/R}}$ is \cite{hesselb}
\begin{equation}
  \sigma_{\tilde{e}_{L/R}} = (f^{L/R}_{e1}f^{L/R}_{e2})^2 
  \left[(1 - P^- P^+) \mp (P^- - P^+) \right]
  \tilde{\sigma}( s,m_{\tilde{e}_{L/R}},m_{\tilde{\chi}^0_{1/2}}) 
\end{equation}
where $P^-$ and $P^+$ are
the longitudinal polarizations of the electron
and positron beam, respectively, and $\tilde{\sigma}$ is a function of
the beam energy and the particle masses.
For selectron masses of the same order of magnitude $m_{\tilde{e}_L}
\approx m_{\tilde{e}_R}$
the polarization behavior is significantly determined
by the ratio
\begin{equation}
r_f=\frac{(f^R_{e1}f^R_{e2})^2}{(f^L_{e1}f^L_{e2})^2}
\end{equation}
of the couplings of the produced neutralinos to
the left handed and right handed electrons and selectrons $f^{L/R}_{ei}$
\cite{hesselb, gudidis}.
The effects of different masses, 
especially the extreme cases $m_{\tilde{e}_L} \ll (\gg)$
$m_{\tilde{e}_R}$, are shortly mentioned in the following section.

In this contribution we consider longitudinal beam polarizations of 
$P^-=\pm 0.85$ for electrons and $P^+=\pm 0.6$ for positrons
\cite{martynOxford}.
The solid lines in Figs.~\ref{mssm}(b), \ref{e6plus}(b) and
\ref{e6plus35}(b)
show the contour for $r_f=1$. 
In the NMSSM and the \es\ model with negative $M'$
(Figs.~\ref{nmssm}(b) and \ref{e6minus}(b)) it is always $r_f>1$.
This contour line
clearly separates the gray shaded regions with maximum cross sections
for polarization
configurations $(+-)$ (right handed polarized electrons and
left handed polarized positrons) or $(-+)$ (vice versa) in the MSSM
(Fig.~\ref{mssm}(b)) in the gaugino region ($M_2 \lesssim 2|\mu|$). In
the \es\ model (Figs.~\ref{e6plus}(b) and \ref{e6plus35}(b))
the situation is more complicated because the higgsino components of
the $\tilde{\chi}^0_1$ are larger than the gaugino components 
in the whole parameter space. Nevertheless the contour $r_f=1$ 
approximately describes the polarization behavior.
The polarization effects in detail are discussed in the following
section.

\section{Polarized cross sections}
We compare the longitudinally polarized cross sections $\sigma^{(+-)}$ and
$\sigma^{(-+)}$ where $(+-)$ and $(-+)$ denotes the polarization
configuration $(P^- = + 0.85,P^+ = - 0.6)$ and $(P^- = - 0.85,P^+ = +
0.6)$, respectively.
These polarization asymmetries may help to
distinguish between the SUSY models.
Figs.~\ref{mssm}(b) -- \ref{e6plus35}(b) present the regions where
the different polarizations $(+-)$ or $(-+)$ lead to
maximum cross sections. Polarization configurations
with both beams polarized in the same direction or only one beam polarized
are always smaller.
 
The following remarks apply to all models:
If the gaugino-character of $\tilde{\chi}^0_1$ and $\tilde{\chi}^0_2$
is not too small (which is the case in the MSSM for $M_2 \lesssim 2 |\mu|$)
$\sigma^{(-+)}$ dominates if the gaugino-coupling of the left
selectron is larger than that of the right one,
$r_f < 1$, and the mass difference
between the selectrons is rather small. Analogously for $r_f > 1$,
$\sigma^{(+-)}$ turns out to be the dominating cross section.

This situation may change for other selectron masses.
If $m_{\tilde{e}_L} \gg (\ll)\; m_{\tilde{e}_R}$ the exchange of the
left (right)
selectron in the neutralino production is strongly suppressed and 
$\sigma^{(+-)}$ ($\sigma^{(-+)}$) dominates in the whole gaugino region,
cf.~\cite{gudilc2000}.

If the gaugino components of the produced neutralinos can be neglected, 
the ratio of the neutralino-selectron-electron couplings plays no role
and the asymmetry in the MSSM and NMSSM
\begin{equation}\label{asyhiggs}
  \frac{\sigma^{(-+)} - \sigma^{(+-)}}{\sigma^{(-+)} + \sigma^{(+-)}}
  \approx \frac{|P^-| + |P^+|}{1 + |P^-| |P^+|}\,
          \frac{L^2 - R^2}{L^2 + R^2}
  > 0
\end{equation}
depends only on the electron-$Z$ couplings $L = -1/2+\sin^2\theta_W$
and $R =  \sin^2\theta_W$.
So $\sigma^{(-+)}$ is generally largest in the MSSM higgsino region
$M_2 \gtrsim 2|\mu|$, whereas
in the NMSSM the gaugino components of $\tilde{\chi}^0_2$
always cause $\sigma^{(+-)}>\sigma^{(-+)}$
for $M_2 < 500$ GeV.
In the \es\ model formula~(\ref{asyhiggs}) becomes more
complicated due to $Z'$ exchange.

Because of $r_f >1$ in the NMSSM and in
the \es\ model with $M'=-50$ TeV $\sigma^{(+-)}$ dominates in
the whole parameter space.
In the \es\ model with positive $M'$ both cases $r_f>1$ and $r_f<1$
occur. Thus $\sigma^{(+-)}$ as well as $\sigma^{(-+)}$ can be
largest. The form of the respective parameter regions crucially
depends on the value of $M'$ (compare Figs.~\ref{e6plus}(b) and
\ref{e6plus35}(b)) and is only approximately given by the contours
$r_f =1$ because here the higgsino character of $\tilde{\chi}^0_1$ is
always rather large.
 
In all models the cross sections for both beams polarized are more
than two times larger than the unpolarized cross sections in a large
fraction of the parameter space. To describe this fact we define the
ratio
\begin{equation}
  r^{(\pm\mp)} = \frac{\sigma^{(\pm\mp)}}{\sigma^{(00)}} \,.
\end{equation}
In the MSSM for  $M_2 \lesssim 2|\mu|$ (gaugino-region) one  nearly
always obtains 
$r^{(+-)} > 2$ or $r^{(-+)} > 2$ in the respective regions. In the
NMSSM and the \es\ model with negative $M'$ it is always $r^{(+-)} >
2$ and even $r^{(+-)} > 2.7$ (NMSSM) or $2.9$ (\es\ model) in a large
fraction of parameter 
space. And also for positive $M'$ $r^{(+-)}>2$ or $r^{(-+)}>2$ holds
in large parameter regions. Thus polarization of both beams is an
important tool to increase the event rates.

Finally we describe the parameter regions 
($0 \le M_2 \le 500$~GeV, $-500~\textrm{GeV} \le \mu,\lambda x \le 500$~GeV)
where significant differences
between the SUSY models arise in neutralino production at a linear
collider if both beams are polarized. In the extended models with a 
light singlino-like neutralino ($m_{\chi_1^0} \approx 30$ GeV) the
polarization configuration $\sigma^{(+-)}$ dominates for
$\lambda x > 0$, while in the MSSM $\sigma^{(-+)}$ is largest apart
from a narrow band between $M_2 \approx \mu$ and $M_2 \approx 2 \mu$.
If the mass of the singlino-like
neutralino, however, increases also the \es\ model with $M'=-35$~TeV
allows $\sigma^{(-+)} > \sigma^{(+-)}$ in some parameter regions.
In the MSSM, NMSSM and \es\ model with negative $M'$ the situation for 
$\mu$ or $\lambda x < 0$ is similar to $\mu,\lambda x > 0$. In the
\es\ model with positive $M'$ the polarization asymmetries strongly
depend on $M'$. For $M'=50$~TeV with a lighter singlino
$\sigma^{(+-)}$ dominates for $M_2 \ll -\lambda x$ and  $M_2 \gg
-\lambda x$ contrary to the MSSM, whereas for $M'=35$~TeV
$\sigma^{(+-)}$ dominates only for large $M_2$.

\section{Conclusion}
We summarize our results in two points:
\begin{itemize}
\item A high luminosity linear collider is sufficient for
the direct production of neutralinos 
with dominant singlino components in extended supersymmetric models. 
Polarization of both beams
is an important tool to increase the event rates by a factor of 2 or 3.
\item The Minimal Supersymmetric Standard Model and extended 
supersymmetric models show a significant different polarization behavior 
for neutralino production in wide regions of the parameter space.
\end{itemize}

\section*{Acknowledgment}
We thank G.~Moortgat-Pick for many valuable discussions.
This work was supported \linebreak
by the Deutsche Forschungsgemeinschaft (DFG) under contract No.\
FR~1064/4-1, \linebreak
by the Bundesministerium f\"ur
Bildung und Forschung (BMBF) under contract No. \linebreak
05~HT9WWA~9 and by
the Fonds zur F\"orderung der wissenschaftlichen Forschung of Austria,
project No.~P13139-PHY.


\begin{thebibliography}{99}

\bibitem{gudi}G. Moortgat-Pick and H. Fraas, Phys. Rev. \textbf{D 59}
  (1999) 015016;\\
  G. Moortgat-Pick, H. Fraas, A. Bartl, and W. Majerotto,
  Eur. Phys. J. \textbf{C 9} (1999) 521; Eur. Phys. J. \textbf{C 9}
  (1999) 549(E). 

\bibitem{sitges} G. Moortgat-Pick, S. Hesselbach, F. Franke, and
  H. Fraas, WUE-ITP-99-023, hep-ph/9909549, contribution to the
  Proceedings of the \emph{4th International Workshop on Linear Colliders
  (LCWS99)}, Sitges, Barcelona, Spain, April 28 -- May 5, 1999. 

\bibitem{nmssm}F. Franke and H. Fraas,
     Z. Phys. \textbf{C 72} (1996) 309 and references therein.

\bibitem{franke} F. Franke, H. Fraas, and A. Bartl, 
     Phys. Lett. \textbf{B 336} (1994) 415.

\bibitem{hr}J.L. Hewett and T.G. Rizzo, Phys. Rep. \textbf{183} (1989)
  193.

\bibitem{e6model}J.F. Gunion, L. Roszkowski, and H.E. Haber,
  Phys. Rev. \textbf{D 38} (1988) 105;\\
  M.M. Boyce, M.A. Doncheski, and H. K\"onig,
  Phys. Rev. \textbf{D 55} (1997) 68;\\
  M. Cveti\v{c}, D.A. Demir, J.R. Espinosa, L. Everett, and P. Langacker,
  Phys. Rev. \textbf{D 56} (1997) 2861;\\
  T. Gherghetta, T.A. Kaeding, and G.L. Kane,
  Phys. Rev. \textbf{D 57} (1998) 3178.

\bibitem{abe}F. Abe {\it et al.} (CDF Collaboration),
  Phys. Rev. Lett. \textbf{79} (1997) 2192.

\bibitem{e6neutralino}J. Ellis, K. Enqvist, D.V. Nanopoulos and
  F. Zwirner, Nucl. Phys. \textbf{B 276} (1986) 14;\\
  S. Nandi, Phys. Lett. \textbf{B 197} (1987) 144;\\
  E. Keith and E. Ma, Phys. Rev. \textbf{D 56} (1997) 7155;\\ 
  D. Suematsu, Phys. Rev. \textbf{D 57} (1998) 1738.

\bibitem{lc97}S. Hesselbach, F. Franke, and H. Fraas,
  in \emph{$e^+e^-$ Linear Colliders:
  Physics and Detector Studies, Part E}, Contributions to the
  Workshops, Frascati, London, Munich, Hamburg, Ed. R.~Settles
  (\mbox{DESY~97-123E}, Hamburg, 1997) p. 479.

\bibitem{hesselb}S. Hesselbach, Ph.D. thesis, University of W\"urzburg, 1999.

\bibitem{decarlosespinosa}B. de Carlos and J.R. Espinosa,
  Phys. Lett. \textbf{B 407} (1997) 12.

\bibitem{lepbounds}
  P. Abreu et al. (DELPHI Collaboration), Phys. Lett. \textbf{B 446}
    (1999) 75;\\
  R. Barate et al. (ALEPH Collaboration), CERN-EP/99-014;\\
  G. Abbiendi et al. (OPAL Collaboration), CERN-EP/99-123,
    hep-ex/9909051;\\
  M. Acciarri et al. (L3 Collaboration), CERN-EP/99-127, hep-ex/9910007.

\bibitem{refszen} S. Ambrosanio, G.A. Blair, and P.M. Zerwas, EFCA-DESY
  Linear Collider Workshop, available via WWW at\\
  \verb|http://www.hep.ph.rhbnc.ac.uk/~blair/susy| .

\bibitem{neutprod}A. Bartl, W. Majerotto, and B. M\"osslacher, 
  in {\it $e^+e^-$ Collisions at 500 GeV: The Physics Potential,
        Part B}, Proceedings of the Workshop,
        Munich, Annecy, Hamburg,
        Ed. P.M.~Zerwas, (DESY~92-123B, Hamburg, 1992) p.~641.

\bibitem{bartlfraasneut}A. Bartl, H. Fraas, W. Majerotto, and N. Oshimo,
  Phys. Rev. \textbf{D 40} (1989) 1594.

\bibitem{lumi}R. Heuer, F. Richard, and P. Zerwas, \emph{Reference
    processes}, available via WWW at\\
  \verb|http://www.desy.de/~schreibr/reference-processes.html| .

\bibitem{gudidis}G. Moortgat-Pick, Ph.D. thesis, University of
  W\"urzburg, 1999.

\bibitem{martynOxford}H.-U. Martyn, talk presented at the Workshop
  \emph{2nd ECFA/DESY Study on Physics and Detectors for a Linear
  Electron-Positron Collider}, Oxford, March 20 -- 23, 1999,
  transparencies available via WWW at\\
  \verb|http://hepnts1.rl.ac.uk/ECFA_DESY_OXFORD/scans/0120_Martyn.pdf| .

\bibitem{gudilc2000}G. Moortgat-Pick, A. Bartl, H. Fraas, and
  W. Majerotto, DESY~00-002, hep-ph/0002253,
  contribution to the Proceedings of the
  \emph{2nd Joint ECFA/DESY Study on 
    Physics and Detectors for a Linear Electron-Positron Collider}.

\end{thebibliography}
\end{document}